\documentclass[runningheads]{llncs}
\usepackage{booktabs} 

\usepackage{array}
\newcolumntype{C}{>{\centering\arraybackslash}m{4.0cm}}
\usepackage{graphicx}
\usepackage{cite}
\usepackage[misc]{ifsym}
\usepackage{listings}
\usepackage{pgfplots}
\pgfplotsset{compat=1.14}
\usepackage{subcaption}
\begin{document}
\title{Visual Summarization of Scholarly Videos using Word Embeddings and Keyphrase Extraction}
\titlerunning{Visual Summarization of Scholarly Videos}

\author{Hang Zhou\inst{1} \and Christian Otto\inst{2}\orcidID{0000-0003-0226-3608}\and Ralph Ewerth\inst{1,2}\orcidID{0000-0003-0918-6297}}

\authorrunning{Zhou et al.}
\institute{L3S Research Center, Leibniz Universit\"at Hannover, Germany\and Leibniz Information Centre for Science and Technology (TIB), Hannover, Germany \newline\email{zhhz417@gmail.com, first.lastname@tib.eu}}
\maketitle             
\begin{abstract}
Effective learning with audiovisual content depends on many factors. Besides the quality of the learning resource's content, it is essential to discover the most relevant and suitable video in order to support the learning process most effectively. Video summarization techniques facilitate this goal by providing a quick overview over the content. It is especially useful for longer recordings such as conference presentations or lectures. In this paper, we present an approach that generates a visual summary of video content based on semantic word embeddings and keyphrase extraction. For this purpose, we exploit video annotations that are automatically generated by speech recognition and video OCR (optical character recognition). 
We demonstrate the feasibility of the proposed approach through its incorporation into the TIB AV portal (http://av.tib.eu/), which is a platform for scientific videos. The accuracy and usefulness of the generated video content visualizations is evaluated in a user study. 

\end{abstract}

\section{Introduction}
\label{sec:introduction}

The massive growth of online video platforms underline the role of audio-visual content as one of the most commonly used sources of information not only for entertainment, but also in learning related scenarios. Exploring a large collection of videos in order to find the most relevant candidate for a specific learning intent can be overwhelming and therefore inefficient. This is especially true for longer videos if the title alone is not able to capture all parts and aspects of the content. Approaches for \textit{video summarization} address this problem by analyzing the visual content and generating an overview by the combination of identified key sequences and frames. However, such approaches struggle with videos, where the visual content lacks variance or is mostly comprised of concepts with low \textit{visualness}~\cite{yanai2005image}, e.g., abstract concepts. Scientific and educational videos often share this characteristic, for example, tutorials or lecture recordings of the STEM subjects (Science, Technology, Engineering, and Mathematics) like chemistry or computer science. 

In this paper, we propose an interactive visualization approach in order to summarize the content of scientific or educational videos. The goal is to provide a tool that facilitates the explorative search capabilities of respective video portals and thus, making learning for the end user more efficient and satisfying. Our approach makes use of automatically extracted video annotations and entities, which significantly enrich the usually available, conventional metadata. These entities are generated from the 1) speech transcript, 2) visual concept classification, and 3) text extracted using optical character recognition (OCR). Such kind of metadata is available for videos of the TIB AV-Portal (https://av.tib.eu), which is run by the Leibniz Information Centre for Science and Technology (TIB). The metadata are provided to the public as open data as well. For these reasons, we choose the TIB AV portal as the basis platform and incorporate the proposed system there. Our system utilizes these data and generates a comprehensive, interactive visualization by combining semantic word embeddings and keyphrase extraction methods. We demonstrate how to display the visualization on the actual website with a \textit{GreaseMonkey} script, which is also a pre-requisite for our user study that investigates the usefulness of the proposed approach for video content visualization. 

The paper is structured as follows: Section 2 discusses the related work for video summarization and other related areas, while Section 3 introduces the different components of our system and the utilized dataset. Section 4 describes the experimental setup and discusses the results. Lastly, section 5 concludes the paper and briefly outlines areas of future work. 
\section{Related Work}
\label{sec:relatedwork}
\subsubsection{Video Summarization}
\label{sec:video_summarization}
The vast majority of video summarization algorithms rely on visual features and are very domain-specific (e.g., movies, sports, news, documentary, surveillance, etc.), resulting in a large number of different approaches. The focus of these approaches can be dominant concepts~\cite{over2008trecvid}, user preferences~\cite{lu2013story}, query context~\cite{wang2012event} or user attention~\cite{hua2005generic}. A typical result of these approaches is a sequence of keyframes or a video excerpt comprising the most important parts of a video.
More recent methods treat video summarization as an optimization problem~\cite{zhang2016summary, gygli2015video, elhamifar2017online} or they utilize recurrent neural networks~\cite{zhang2016video, zhao2018hsa} based on, for instance, long short-term memory cells (LSTMs), which are able to capture temporal or sequential information very well. Another use case for LSTMs is proposed by Mahasseni et al.~\cite{mahasseni2017unsupervised}, who suggest a generative adversarial network (GAN) consisting of an LSTM-based autoencoder and a discriminator.
There are also methods that include textual information (e.g., tags~\cite{hong2011beyond} or full documents~\cite{li2016multimedia}) which result in a storyboard that provides short titles for each keyshot. This is in particular useful for news summarization. Scientific or scholarly videos provide a greater challenge in this respect, since their visual content often lacks visualness. Consequently, summarization techniques focus even more on textual metadata. Chang et al.~\cite{chang2011keyword} combine image processing, text summarization and keyword extraction techniques resulting in a multimodal surrogate. A word cloud is generated, where more important words are displayed with a bigger font size, and also a set of three to four thumbnails with a short transcription. 

In this paper, we go one step further and show how to summarize the content solely based on textual information. The core techniques to create a video summarization utilized in this paper are keyphrase extraction and measures for semantic text similarity. The related work in these respective areas is subsequently described.
\subsubsection{Keyphrase Extraction}
\label{sec:keyphrase}
Hasan and Ng~\cite{hasan2014automatic} describe that keyphrase extraction techniques are generally comprised of two steps. First, a list of possible candidate phrases is identified, and then these candidates are ranked according to their importance. This is realized by a wide range of approaches that can be categorized into supervised and unsupervised methods. 
Early supervised algorithms rely on, for instance, decision trees~\cite{turney2000learning}. Hulth~\cite{hulth2004reducing} extend this approach by adding linguistic features to a bagged decision tree classifier while also extending previous work by filtering incorrectly assigned keywords with different feature pairs. Another approach~\cite{ercan2007using} utilizes lexical chains based on a WordNet ontology, which is associated with features such as first occurrence position, last occurrence position, and word frequency. Additionally, support vector machines~\cite{wang2005keyphrases}, maximum entropy classifiers~\cite{kim2009re, liu2011supervised}, conditional random field models~\cite{zhang2008automatic}, logistic regression~\cite{haddoud2014accurate} and neural networks~\cite{wang2006automatic, jo2003neural} have been used to solve the task of finding the most important phrases in a document. 

All of the aforementioned techniques share the drawback, that the training data require manual labeling, which generally introduces unreal experimental data and is time-consuming and resource-intensive. Thus, unsupervised approaches moved into the focus of attention. Their task is to automatically discover the underlying structure of a dataset without human-labeled keyphrases. To summarize, the two most popular methods are \textit{graph-based ranking} and \textit{topic-based clustering}. The idea behind graph-based algorithms is to construct a graph of phrases, which are connected with weighted edges that describe their relation derived from the frequency of their co-occurrence~\cite{mihalcea2004textrank}. Topic-based clustering methods use statistical language models, which contain the probability of all possible sequences of words~\cite{blei2003latent}.
Recently, fusions of these two directions gain attention, namely TopicRank~\cite{bougouin2013topicrank}, PositionRank~\cite{florescu2017positionrank} and MultiPartiteRank~\cite{boudin2018unsupervised}. The latter one, which is also used in our approach, first builds a graph representation of the document and then ranks each keyphrase with a relevance score. In addition, in an intermediate step edge weights are adjusted to capture information about the word's position in the document.

\subsubsection{Semantic Text Similarity}
\label{sec:textsim}
Corpus-based similarity algorithms determine the semantic relation between two textual phrases based on information learned from large corpora like Wikipedia. Particulary neural network approaches benefit greatly from huge amounts of data, leading to the current success of methods such as Word2Vec~\cite{mikolov2013distributed}, GloVe~\cite{pennington2014glove}, and fastText~\cite{bojanowski2017enriching}. They all create word-vector spaces that cover a desired vocabulary size and embed semantically similar words close to one another, while they also allow for mathematical operations on these vectors to unveil relationships. For instance, when applying fastText, the difference vectors between Paris - France and Rome - Italy are almost identical, indicating the vectors describe the relation "capital". If this vector is added to Poland, it therefore leads to Warsaw.
\vspace{-0.3cm}
\section{Visual Summarization of Scientific Video Content}
\label{sec:methodology}
\begin{figure}[htbp]
	\centering
	\includegraphics[width=0.7\textwidth]{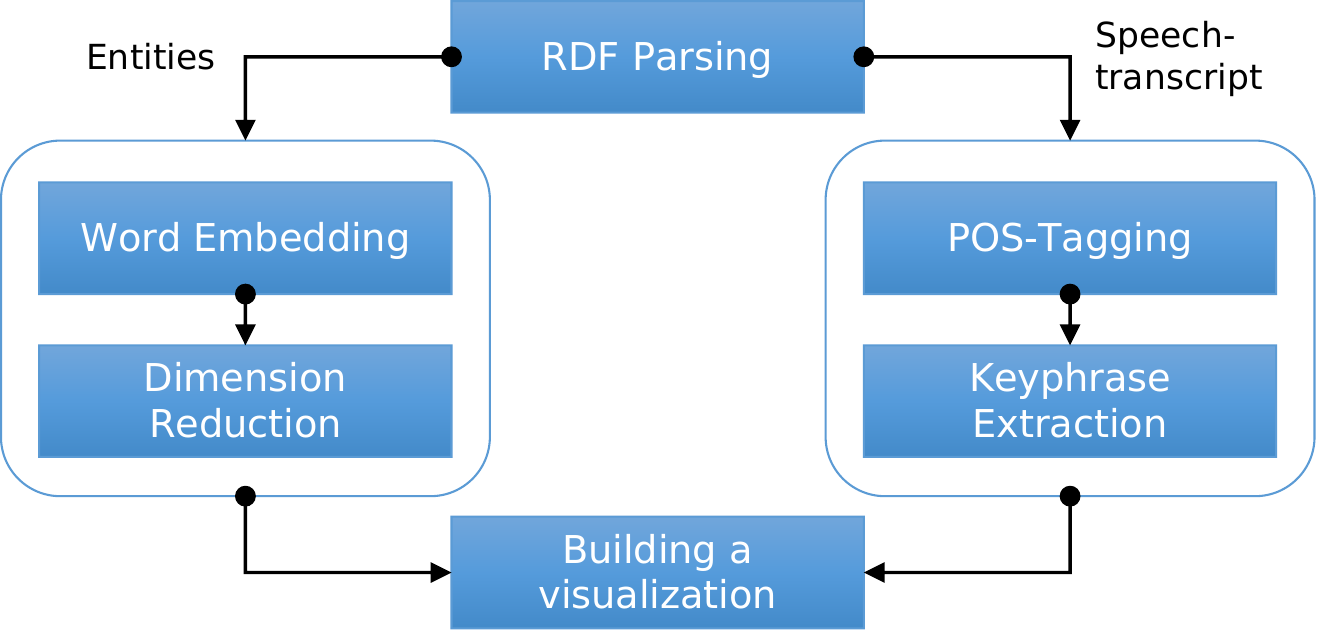}
	\caption{Workflow diagram of the proposed visualization approach.}
	\label{fig:workflow}
	\vspace{-0.3cm}
\end{figure}

In this section, we describe our approach for video content summarization solely based on textual information. The necessary process to summarize a scientific video and display this information in an efficient way~\ref{fig:workflow} consists of four steps: 1) pre-processing, 2) semantic embedding of content related information to generate a bubble diagram, 3) creation of a keyphrase table from the speech transcript, and 4) combining diagram and table to form a visualization. The utilized video dataset from the TIB AV-Portal is available at \url{https://av.tib.eu/opendata}, including the associated metadata as Resource Description Framework (RDF) triples (under Creative Commons License CC0 1.0 Universal). 

\subsubsection{Preprocessing}

To build the RDF graph we use Python 3.6 and the \textit{rdflib} library. Next, we use the query language SPARQL to select videos that contain automatically extracted metadata (this applies only for videos related to the six core subjects of the TIB\footnote{engineering, architecture, chemistry, computer science, mathematics, physics}). An exemplary query can be seen in Listing~\ref{lst:sparql}. 

\lstset{
language=SPARQL, 
basicstyle=\ttfamily\footnotesize,
caption={SPARQL-query that returns all videos which contain automatically analyzed speech transcripts (ASR) and recognized entities.},
captionpos=b,
frame=single,
numbers=left,
label=lst:sparql}
\begin{lstlisting}
PREFIX dcterms: <http://purl.org/dc/terms/>
PREFIX oa:      <http://w3.org/ns/oa#>

SELECT DISTINCT ?url
WHERE {
    ?annotation oa:annotatedBy asr_link.
    ?annotation oa:hasTarget ?videofragment .
    ?videofragment dcterms:isPartOf ?url .}
    
\end{lstlisting}This yields a list of $1756$ videos from multiple languages, which we then query further for the embedded metadata, in particular the \textbf{key entities} which are the result of visual concept classification, optical character recognition and automated speech recognition. Additionally, we crawl the unfiltered speech transcript from the website using the \textit{BeautifulSoup} library.

\subsubsection{Semantic Embedding of Key Entities}
We use fastText to generate word embeddings from the extracted key entities. fastText's tri-gram technique embeds words by their substrings instead of the whole word, for instance, the word \textit{google} would be decomposed into the following tri-grams: {$<$go, goo, oog, ogl, gle, le$>$}. This is a valuable feature for multiple reasons. First, it enables the system to encode misspelled or unknown words. Secondly, the quality of embeddings of the generally longer or compound words of the German language is improved, too. We use the pre-trained model for German language~\footnote{\url{https://fasttext.cc/docs/en/crawl-vectors.html}}, which contains the vocabulary of the German Wikipedia and encodes each word $w$ in a 300-dimensional vector $f_w$.
The visualization of the embedded feature vectors requires dimension reduction to project data onto a two-dimensional space. We apply a linear algorithm (principal component analysis) instead of a non-linear one like t-SNE, since we intend to keep the semantic arrangement laid out by fastText and refrain from clustering the keywords further.

\subsubsection{Keyphrase Extraction}
Input for the keyphrase extraction process is the unfiltered speech transcript, which is already divided into time segments in TIB's AV portal. The required format of the textual information is given by the \textit{pke} toolkit~\cite{boudin:2016:COLINGDEMO} which is shown in Listing~\ref{lst:preptext}. Requirements are a tokenization and part-of-speech tagging (POS tagging), that is the assignment of lexical categories such as noun, verb, adjective, adverb and so on. For this process we use the Python \textit{Natural Language Toolkit (NLTK)}, in particular, the Stanford POS-Tagger which also comes with a pre-trained model for the German anguage\footnote{\url{https://nlp.stanford.edu/software/tagger.shtml}}.
\lstset{
caption={POS-Tagged speech transcript labeled with lexical categories.}, 
captionpos=b,
frame=single,
numbers=none,
label=lst:preptext}
\begin{lstlisting}
wenig/PRON Speicher/NOUN es/PRON kommen/VERB [...]
\end{lstlisting}
Results of the POS tagging process are then passed to the Multipartite Rank~\cite{boudin2018unsupervised} algorithm of the \textit{pke} library in order to perform keyphrase extraction. As stated in Section \ref{sec:relatedwork}, this technique models topics and phrases in a single graph, and their mutual reinforcement together with a specific mechanism to select the most import keyphrases are used to generate candidate rankings. We only consider nouns, adjectives, personal pronouns, and verbs ('NOUN', 'ADJ', 'PROPN', 'VERB') and dismiss all words given by \textit{NLTK}'s collection of German stop words. The remaining parameters are $alpha=1$, which controls the weight adjustment mechanism, and the $threshold=0.4$ for the minimum similarity for clustering (default: $0.25$). We decide to set this value to $0.4$ due to the high similarity of topics in a single video. The linkage method was set to \textit{average}. Finally, we choose to retrieve the $20$ highest ranked keyphrases of every time stamp for our keyphrase table that will become part of the visualization.

\subsubsection{Visualization of Results}

Finally, we display the recognized, embedded entities in an interactive graph with the properties shown in Table~\ref{tab:properties} and combine it with the keyphrase table generated in the last section.

\begin{table}
	\begin{tabular}{| C | C | C |}
		\hline
		\textbf{Components}  & \textbf{Meaning}       & \textbf{Approach}              \\
		\hline \hline
		circle      & key topics    & recognized entities   \\ 
		\hline
		the size of a circle & importance of the topic & the frequency of the entity \\
		\hline
		arrangement & similarity between topics & word embeddings\\
		\hline
		table       & timestamp based summary of the speech transcript & keyphrase extraction \\
		\hline
	\end{tabular}
    \centering
    \vspace{0.1cm}
	\caption{Overview over the properties of the visualization.}
	\label{tab:properties}
\vspace{-0.2cm}
\end{table}
We choose a bubble diagram as opposed to Chang et al.'s~\cite{chang2011keyword} word cloud. This allows us to illustrate and emphasize also on the distance, between related or unrelated keywords, which reflects (dis)similarity. In addition, smaller differences in area sizes are visually easier to perceive than font sizes. We decided against other alternative implementations such as TextArc~\cite{paley2002textarc} since we aimed for a more intuitive approach. The inclusion of the temporal dimension using ThemeRiver~\cite{havre2002themeriver} did not deliver consistent results for videos that were short, or contained only few keywords. In addition, ThemeRiver is less suitable to represent the similarity of several entities.

\begin{figure}[htbp]
	\centering\fbox{
	\includegraphics[width=0.8\textwidth]{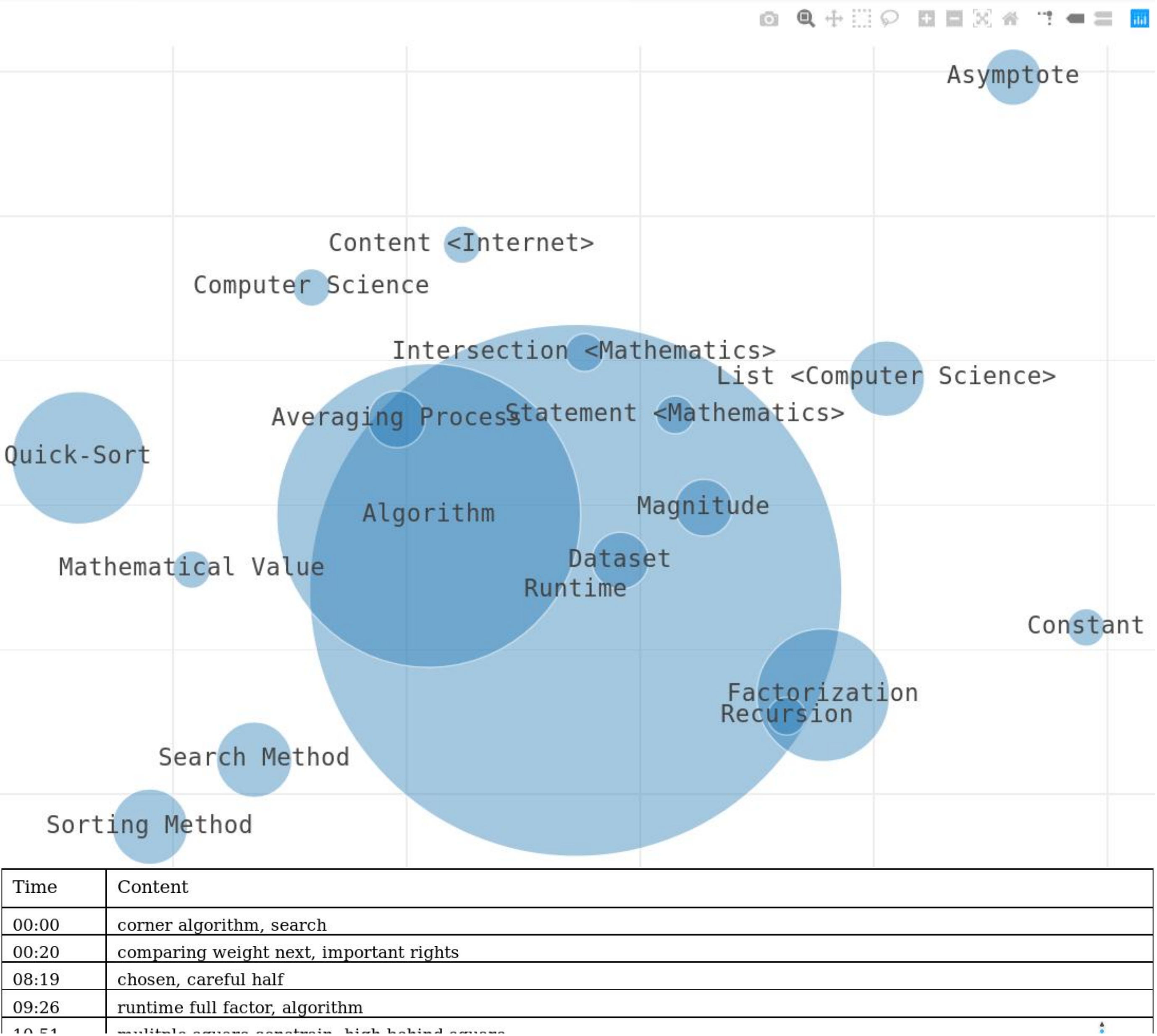}}
	\caption{Visualization of video \url{https://av.tib.eu/media/9557} titled "Bubblesort, Quicksort, Runtime" incorporated via GreaseMonkey in the live website as portrayed during the user study comprised of the visualization itself, a toolbar and the keyphrase table. Note: Translated for better comprehensibility.}
	\label{fig:vis1}
\end{figure}

\begin{figure}[htbp]
	\centering\fbox{
	\includegraphics[width=0.5\textwidth]{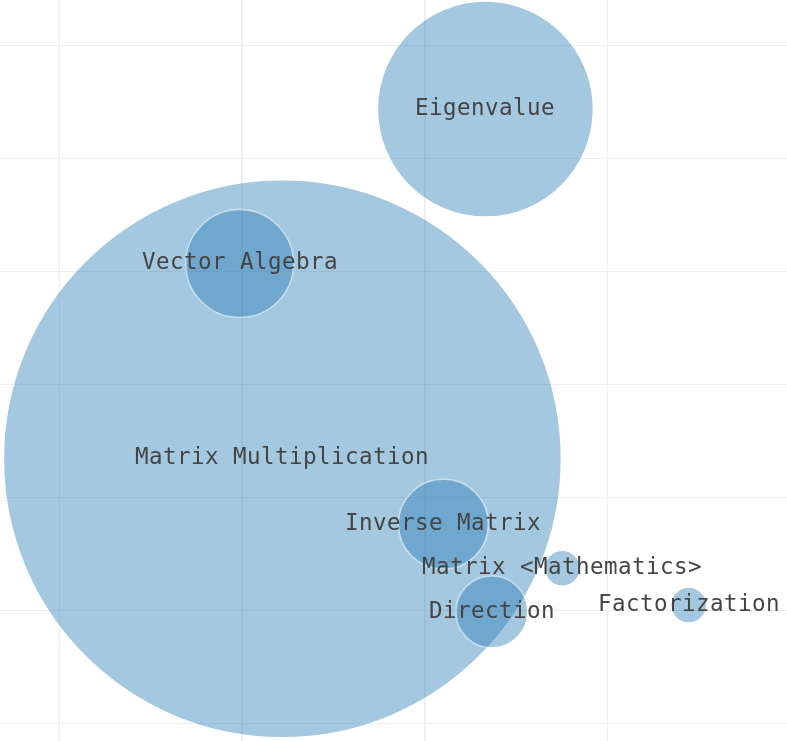}}
	\caption{Visualization of video \url{https://av.tib.eu/media/10234} titled "Eigenwerte, Eigenvektoren" (eng: "eigenvalues, eigenvectors"). Note: Entities were translated for better comprehensibility.}
	\label{fig:vis2}
\end{figure}

The actual implementation is done in Javascript and the \textit{Plot.ly} API\footnote{\url{https://plot.ly/api/}}. As displayed in Figures~ \ref{fig:vis1} and \ref{fig:vis2}, the visualization is comprised of circles of different sizes each representing a topic and its importance. An interactive toolbar is displayed on the upper right allowing the user to explore the graph easily. At the bottom, a keyphrase table indicates the main topic of each time segment.
\vspace{-0.3cm}
\section{Experiments and Results}
\label{sec:experiments}

We conducted a user study to evaluate the quality and usefulness of the proposed visualization approach. 10 participants were recruited, of which 8 were male and 2 female. Their ages were between 21 and 30, and their educational levels between high-school and master. Seven participants study computer science, one mechanical engineering and one mathematics. All of them are fluent in German, four of them are native speakers.
\textbf{Task I} of the study investigates how precisely the visual summary represent the video content. Therefore, 10 videos with a duration of 5 to 30 minutes were randomly assigned to each participant. Then, the user had to rate how well the presented visualization matches the video content, based on the following options: "0" - no correlation, "1" - slight match, "2" - good match, "3" - exact match. 
\textbf{Task II} aimed to evaluate if the visualization is a useful tool to provide a quick overview of the video content, or if it is no improvement over the current state of the website. The participants could choose one of the following options to rate the usefulness: "0" - not helpful at all, "1" - slightly helpful, "2" - moderately helpful, "3" - very helpful, "4" - extremely helpful, and had to give a short statement about their reasoning. Figure~\ref{fig:result1} shows the distribution of the 100 gathered ratings, while Figure~\ref{fig:result2} shows the results of Task II.

\begin{figure}[!ht]
\centering
\begin{subfigure}{0.45\textwidth}
\begin{tikzpicture}
    \begin{axis}[
    title={Accuracy},
    ybar=0pt,
    ymin=0,
    width=\textwidth,
    height=0.25\textheight,
    ytick={0,10,20,30,40,50},
    bar width=12pt,
    symbolic x coords={0, 1, 2, 3},
    ylabel={\# of videos},
    xtick=data,
    ymajorgrids=true]
        \addplot[fill=red] coordinates {
        (0, 6)
        (1, 26)
        (2, 42)
        (3, 26)
        };
    \end{axis}
\end{tikzpicture}
\caption{Results of Task I of the user study evaluating the correlation of the visualization to the video content. From "0" - no correlation to "3" - exact match.}
\label{fig:result1}
\end{subfigure}%
\hfill
\begin{subfigure}{0.45\textwidth}
    \begin{tikzpicture}
        \begin{axis}[
        title={Helpfulness},
        ybar=0pt,
        ymin=0,
        width=\textwidth,
        height=0.25\textheight,
        ytick={0,1,2,3,4,5},
        bar width=12pt,
        symbolic x coords={0, 1, 2, 3, 4},
        ytick={0,1,2,3,4},
        ylabel={\# of participants},
        xtick=data,
        ymajorgrids=true]
            \addplot[fill=blue] coordinates {
            (0, 0)
            (1, 1)
            (2, 4)
            (3, 3)
            (4, 1)
            };
        \end{axis}
    \end{tikzpicture}
    \caption{Results of Task II of the user study showing the perceived helpfulness of the visualization. From "0" - not helpful at all to "4" - extremely helpful.}
\label{fig:result2}
\end{subfigure}%
\label{tab:results}
\end{figure}
\subsubsection{Discussion} 
Figure~\ref{fig:result1} shows that $68\%$ of the visualizations were good or better, while $26\%$ only provided a slight match or did not correlate at all to the video content ($6\%$).
Positive examples, as can be seen in Figure 2, successfully provide the user with a summarization of the video content. The first example, video 9557, explains the runtime behavior of the sorting algorithms Bubblesort and Quicksort. The largest circle in the visualization is Runtime ("Laufzeit") and represents well the main topic. Also related topics from computer science covered in the video like sorting methods ("Sortierverfahren"), algorithm ("Algorithmus") and Quick-Sort itself are closely arranged on the left, while related topics from mathematics like factorization ("Faktorisierung"), asymptote ("Asymptote") and statement ("Aussage <Mathematik>") are grouped on the right. The second positive example (video 10234), which talks about eigenvalues and eigenvectors, is mainly represented by the entity matrix multiplication ("Matrizenmultiplikation") and vector ("Vektor"), but also shows more detailed aspects of that topic, namely vector algebra ("Vektorrechnung"), inverse matrix, gradient and of course, eigenvector and eigenvalue. 

The results of the keyphrase extraction, as can be seen in Figure~\ref{fig:vis1}, were less helpful. The main reason is most likely the nature of the automatic speech transcripts, which usually differ from written text. They often consist of incomplete sentences, misspelled words, missing punctuation, and falsely recognized  words which can change interpretations of a sentence completely. Since common keyphrase models are suited for proper textual content, there is still room for improvement in our scenario.

\begin{figure}[htbp]
	\centering\fbox{
	\includegraphics[width=0.7\textwidth]{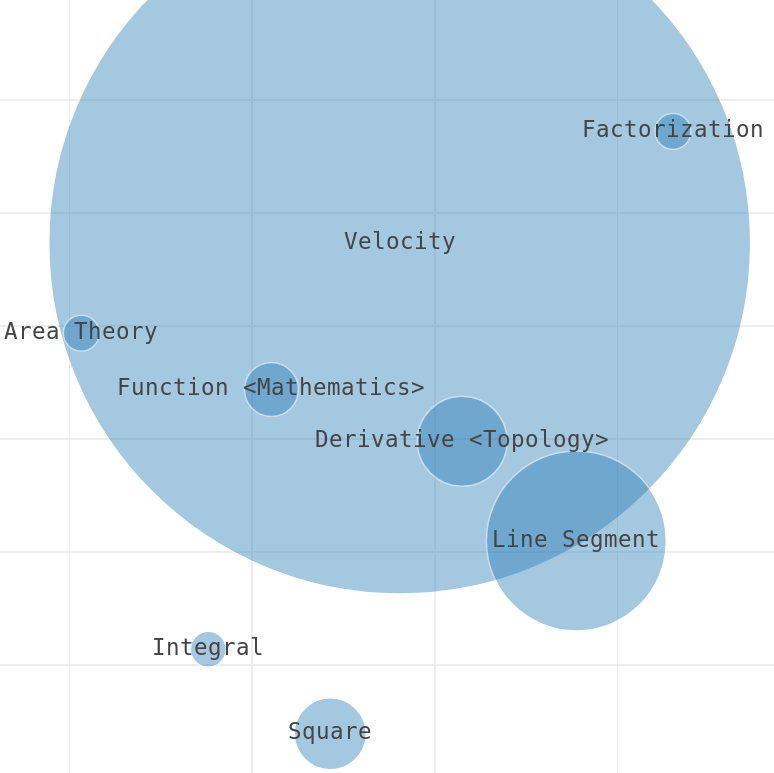}}
	\caption{Visualization of video \url{https://av.tib.eu/media/9915}. It demonstrates the effect of very common entity "Geschwindigkeit" (eng: velocity), which was used frequently by the speaker during an example scenario, but is misleading since the video talks about arc length computation. Note: Entities were translated for better comprehensibility.}
	\label{fig:badvis}
\end{figure}

In order to find out what led participants to give the rating "uncorrelated", we reviewed these 6 videos and found that they came from the subject engineering and had very application-specific content, which might be a limitation of the system. One video, for instance, discusses the cause, consequences and solutions of driftwood accumulation on bridges leading to overflowing rivers (\url{https://av.tib.eu/media/11442}). A lot of technical terms, switching contexts from real world, to model testing to technical considerations paired with topic specific phrases yielded a visualization which was only marginally helpful. 
Finally, the reason that more results present a "good match" instead of an exact match is most likely due to the nature of the entities extracted from the speech transcript. For example, videos and tutorials from the field of mathematics contain a lot of terms that are important when explaining a concept, but are rather general and not closely related to the topic itself. That includes words like "square", "point" and "integral". Yet, these words are captured by the system and are present in the dataset, but they contribute only marginally to the comprehension of the video even though they appear very frequently. A respective less useful summary result is exemplary presented in Figure~\ref{fig:badvis}. 
This circumstance is also reflected in the results of Task II, where our participants agreed that the visualization would be more helpful if the redundant keywords were omitted.

\section{Conclusion and Future Work}
\label{sec:conclusion}
In this paper, we have presented a system that summarizes and displays the content of scholarly videos in order to support semantic search in video portals. Based on entirely automatic video content analysis as conducted in the TIB AV Portal, we have proposed an approach that leverages the resulting metadata and generates an interactive visualization and a keyphrase table to outline the content of a video. Different techniques like POS-Tagging, semantic word embeddings and keyphrase extractions were exploited in our approach. The usefulness of the visualization was evaluated in a user study that demonstrated the feasibility of the proposed visual summarization, but also indicated areas for future work. For instance, we plan to implement reliable filters for keywords that are not closely related to the content to provide a better user experience. 

\bibliographystyle{splncs04.bst}
\bibliography{references}

\end{document}